\documentclass[amsmath,amssymb,
onecolumn,apm]{revtex4-2}

\usepackage{graphicx}

\begin{document}

\title{The accelerated expansion in 
$F(G,T_{\mu \nu}T^{\mu \nu})$ gravity
}

\author{Mihai Marciu}
\email{mihai.marciu@drd.unibuc.ro}
 \affiliation{Faculty of Physics, University of Bucharest, Bucharest-Magurele, Romania}
\author{Dana Maria Ioan}%
 \email{idana91@yahoo.com}
\affiliation{ 
Faculty of Physics, University of Bucharest, Bucharest-Magurele, Romania
}%

\date{\today}

\begin{abstract}
In the present manuscript the basic Einstein--Hilbert cosmological model is extended, by adding a new functional $F(G, T_{\mu\nu}T^{\mu\nu})$ in the fundamental action, encoding specific geometrical effects due to a nontrivial coupling with the Gauss-Bonnet invariant ($G$), and the energy--momentum squared term ($T_{\mu\nu}T^{\mu\nu}$). After obtaining the corresponding gravitational field equations for the specific decomposition where $F(G, T_{\mu\nu}T^{\mu\nu})=f(G)+g(T_{\mu\nu}T^{\mu\nu})$, we have explored the physical features of the cosmological model by considering the linear stability theory, an important analytical tool in the cosmological theory which can reveal the dynamical characteristics of the phase space. The analytical exploration of the corresponding phase space structure revealed that the present model can represent a viable dark energy model, with various stationary points where the effective equation of state corresponds to a de--Sitter epoch, possible explaining the early and late time acceleration of the Universe. 
\end{abstract}

\maketitle

\section{Introduction}
\label{intro}

\par 
In the present cosmological context the accelerated expansion \cite{Frieman:2008sn} represents an enigmatic phenomenon associated with the evolution of our Universe at the level of background dynamics. This phenomenon has been discovered almost two decades ago \cite{Li:2012dt, Joyce:2016vqv}, triggering various developments in science and technology. The simplest dark energy model explaining the accelerated expansion of our Universe is associated to the $\Lambda$CDM model \cite{Bamba:2012cp, Copeland:2006wr,Peebles:2002gy, Padmanabhan:2002ji}, a specific cosmological theory which is based on a cosmological constant $\Lambda$ added to the Einstein's field equations. The $\Lambda$CDM model suffers from various theoretical limitations \cite{Moore:1994yx, 10.1111/j.1745-3933.2011.01074.x, Perivolaropoulos:2021jda} and cannot explain the dynamical evolution of the dark energy equation of state, as probed through various astrophysical observations \cite{Planck:2018vyg, WMAP:2003elm,BOSS:2016wmc, SupernovaSearchTeam:2003cyd, SupernovaCosmologyProject:2011ycw}. In principle the $\Lambda$CDM model \cite{Perivolaropoulos:2021jda} can be regarded as an effective approximate approach which is associated with a constant equation of state, without addressing the $H_0$ tension \cite{Escamilla-Rivera:2022mkc, Poulin:2018cxd, Efstathiou:2013via} in a fundamental manner.
\par 
In the cosmological theories the modified gravity approaches \cite{Nojiri:2017ncd, Nojiri:2006ri, Capozziello:2011et, Tsujikawa:2010zza} represent a novel paradigm which further extends the fundamental action, embedding various invariant components, aiming for a more complete and consistent theory of gravitation. The most natural extension of gravity is represented by the $f(R)$ theory, a specific approach based on a functional which depends on scalar curvature \cite{DeFelice:2010aj}. Since then, many alternative theories have been proposed \cite{Bamba:2012cp, Cai:2015emx, Bahamonde:2015zma, Bahamonde:2021gfp}, aiming for a more consistent theory \cite{Koyama:2015vza, Bull:2015stt,Bahamonde:2020vfj} which can explain the accelerated expansion of our Universe, embedding various dynamical effects associated to the dark sector. In these theories the interplay between matter and geometry has been questioned in different approaches \cite{Harko:2011kv, Harko:2010mv, Haghani:2013oma, Jaybhaye:2023lgr}.
\par 
In the modified gravity theories a particular extension is related to the energy--momentum squared gravity \cite{Roshan:2016mbt, Katirci:2013okf, Board:2017ign}, a novel theory which can explain various physical effects at cosmological scales \cite{Bahamonde:2019urw, Kazemi:2020hep, Chen:2019dip}. The latter theory is constructed by considering the interplay between matter and geometry, taking into account an invariant which is based on a specific self--contraction of the energy--momentum tensor \cite{Roshan:2016mbt}. The energy--momentum squared gravity has attracted some attention in modern cosmological theories \cite{Sharif:2023hbc, Sharif:2023uac, Sharif:2023uyv, Fazlollahi:2023cgp, Yousaf:2023wdw, Akarsu:2023nyl, Sharif:2022akn, Tangphati:2022acb, Nari:2018aqs, Sharif:2022umv, Sharif:2022dzl, Sharif:2022zmw, Sharif:2022oia}, representing a viable approach also from the astrophysical point of view \cite{Akarsu:2018aro, Akarsu:2018zxl}. Specific wormholes solutions have been considered in the energy--momentum squared gravity \cite{Sharif:2021gdv}, analyzing the physical implications. The inclusion of the Gauss--Bonnet topological invariant in the energy--momentum squared gravity has been considered recently \cite{Yousaf:2021xex,Yousaf:2022bwc, Yousaf:2023wdw, Yousaf:2023dqr} for specific relativistic systems.  
\par 
An important approach in modern cosmological theories is related to the Gauss--Bonnet invariant, a special topological component in the four--dimensional space--time \cite{Li:2007jm, Bamba:2010wfw}. The inclusion of the Gauss--Bonnet invariant has been considered in various modern theories of gravitation \cite{Elizalde:2010jx,Shah:2022mme, DeFelice:2009aj,delaCruz-Dombriz:2011oii, delaCruz-Dombriz:2011oii, delaCruz-Dombriz:2011oii, Ganiou:2022sih, Nojiri:2021mxf, PhysRevD.99.043508}, representing a viable approach for specific physical systems, possible explaining the dark energy phenomenon \cite{DeFelice:2008wz}.

\par 
In this paper we shall consider a modified gravity model build in the fundamental framework of the Einstein--Hilbert action, embedding the geometrical interplay between the Gauss--Bonnet invariant and the energy--momentum--squared component \cite{Yousaf:2021xex,Yousaf:2022bwc, Yousaf:2023wdw, Yousaf:2023dqr}. The fundamental action in our model contains a generic functional which depends on the Gauss--Bonnet invariant \cite{delaCruz-Dombriz:2011oii} and the energy--momentum--squared term in a decomposed manner. The physical characteristics are evaluated by considering the linear stability theory for an exponential decomposition, analyzing the phase space structure and the possibility of reaching the accelerated expansion. Such an approach further extends the Einstein--Hilbert action by taking into account the effects due to the geometrical characteristics of space--time, including also the interplay with the matter sector, embedding the elementary properties of the latter component.
\par 
The plan of our paper is the following. In Sec.~\ref{sec:1} we propose the fundamental action for our toy model, obtaining the corresponding modified Friedmann equations. Then, in Sec.~\ref{sec:2} we discuss the physical properties and the emergence of the accelerated expansion in the current cosmological model by considering the linear stability theory in the case of an exponential behavior. Lastly, in Sec.~\ref{sec:3} we summarize the principal obtained results and give the main concluding remarks.

\section{The action and the field equations}
\label{sec:1}

\par 
In what follows we shall propose a cosmological model described by the following action \cite{Yousaf:2021xex,Yousaf:2022bwc, Yousaf:2023wdw}:
\begin{equation}
\label{actiune0}
S=\int d^{4}x\sqrt{-\Tilde{g}} \Bigg[\frac{R}{2} +F(G, T^2)\Bigg]+\int d^4x \sqrt{-\Tilde{g}} L_m, 
\end{equation}
where the generic function embedded into the Einstein--Hilbert action can be decomposed in two specific terms, $F(G, T^2)=f(G)+g(T^2)$. In this case we have assumed a non--linear dependence of the action by the Gauss--Bonnet invariant $(G)$, and the energy--momentum squared invariant $(T^2=T_{\mu \nu}T^{\mu \nu})$ \cite{Yousaf:2021xex,Yousaf:2022bwc}. Before proceeding to the computations of the modified Friedmann relations, we have to specify that the background dynamics can be described by the Robertson--Walker metric: 
\begin{equation}
\label{metric}
ds^2=-dt^2+a^2(t)(dx^2+dy^2+dz^2),
\end{equation}
where $a(t)$ is the cosmic scale factor which characterizes the expansion of the Universe at the large scale structure. In general, the Gauss--Bonnet invariant is defined in the following way, 
\begin{equation}
G=R^2-4R_{\mu\nu}R^{\mu\nu}+R_{\mu\nu\xi\sigma}R^{\mu\nu\xi\sigma}.
\end{equation}
For the above metric \eqref{metric} the Gauss--Bonnet invariant reduces to the following expression  \cite{delaCruz-Dombriz:2011oii},
\begin{equation}
   G=24H^2(H^2+\dot{H}),
\end{equation}
where $H(t)$ represents the Hubble parameter defined as: $H=\frac{\dot{a}}{a}$, where the dot represents the derivative with respect to the cosmic time. In this case the scalar curvature acquires the following expression, 
\begin{equation}
R=6(2H^2+\dot{H}).
\end{equation}
\par 
The second invariant in our action \eqref{actiune0} is represented by the energy--momentum--squared term \cite{Bahamonde:2019urw}, defined as:
\begin{equation}
T^2= T_{\mu\nu}T^{\mu\nu},
\end{equation}
where $T_{\mu\nu}$ describes the energy--momentum of the matter sector, $T_{\mu\nu}=diag[\rho, p, p, p]$, with $\rho$ the density and $p$ the pressure of the matter component which behaves closely as a non--relativistic fluid with zero pressure. If we further assume a barotropic equation of state for the matter sector,
\begin{equation}
    p=w \rho,
\end{equation}
with $w$ the barotropic constant parameter, then the energy--momentum--squared invariant takes the following expression:
\begin{equation}
    T^2=\rho^2(1+3 w^2).
\end{equation}
\par 
The variation of the action \eqref{actiune0} with respect to the inverse metric $g^{\mu \nu}$ lead to the following modified Friedmann relations \cite{Yousaf:2023wdw,Bahamonde:2019urw}:

\begin{equation}
\label{frcon}
3H^2=\rho+\Bigg[G f'(G)-f(G)-24H^3\dot{f'(G)} \Bigg]  
+2\frac{\partial g(T^2)}{\partial T^2} (\rho^2+4\rho p+3p^2) -g(T^2) ,
\end{equation}

\begin{equation}
-3H^2-2 \dot{H}=p+\Bigg[ f(G)- G f'(G) +16 (\dot{H}+H^2) \dot{f'(G)}
\\
+8 H^2 \ddot{f'(G)} \Bigg] + g(T^2),
\end{equation}
where we have assumed the following definitions:
\begin{equation}
    '=\frac{d}{dG},
\end{equation}
\begin{equation}
    \dot{}=\frac{d}{dt},
\end{equation}
\begin{equation}
    \ddot{}=\frac{d^2}{dt^2}.
\end{equation}
\par 
We can further define the energy density associated to the geometrical dark energy component \cite{Bahamonde:2019urw},
\begin{equation}
\rho_{de}=G f'(G)-f(G)-24H^3\dot{f'(G)}   
+2\frac{\partial g(T^2)}{\partial T^2} (\rho^2+4\rho p+3p^2) -g(T^2) ,
\end{equation}
and the corresponding pressure:
\begin{equation}
p_{de}= f(G)- G f'(G) +16 (\dot{H}+H^2) \dot{f'(G)}
\\
+8 H^2 \ddot{f'(G)} + g(T^2).
\end{equation}
\par 
Then, we can define the dark energy equation of state due to the geometrical coupling of the invariant constituents, 
\begin{equation}
    w_{de}=\frac{p_{de}}{\rho_{de}},
\end{equation}
and the total/effective equation of state for the background dynamics:
\begin{equation}
    w_{tot}=-1-\frac{2}{3}\frac{\dot{H}}{H^2}.
\end{equation}
\par 
Lastly, we define the matter density parameter in the usual manner, 
\begin{equation}
    \Omega_m=\frac{\rho}{3H^2}
\end{equation}
and the geometrical dark energy density parameter,
\begin{equation}
    \Omega_{de}=\frac{\rho_{de}}{3H^2},
\end{equation}
satisfying the constraint equation: $\Omega_m+\Omega_{de}=1$.

\section{Dynamical properties for an exponential model}
\label{sec:2}
\par 
In order to study the dynamical properties for the exponential model where $F(G, T_{\mu\nu}T^{\mu\nu})=f_0 e^{\alpha G}+g_0 e^{\beta T^2}$, with $\alpha, \beta, f_0, g_0$ constant parameters, we need to introduce the following auxiliary variables:

\begin{equation}
    s=\Omega_m=\frac{\rho}{3 H^2},
\end{equation}

\begin{equation}
    x=\frac{G}{3 H^2}\frac{df(G)}{dG},
\end{equation}

\begin{equation}
    y=8 H \dot{f'(G)},
\end{equation}

\begin{equation}
    z=2\frac{dg(T^2)}{d(T^2)}\rho (1+4w+3w^2),
\end{equation}

\begin{equation}
    u=\frac{f(G)}{3H^2},
\end{equation}

\begin{equation}
    v=\frac{g(T^2)}{3H^2}.
\end{equation}

\begin{figure}[t]
  \includegraphics[width=6cm]{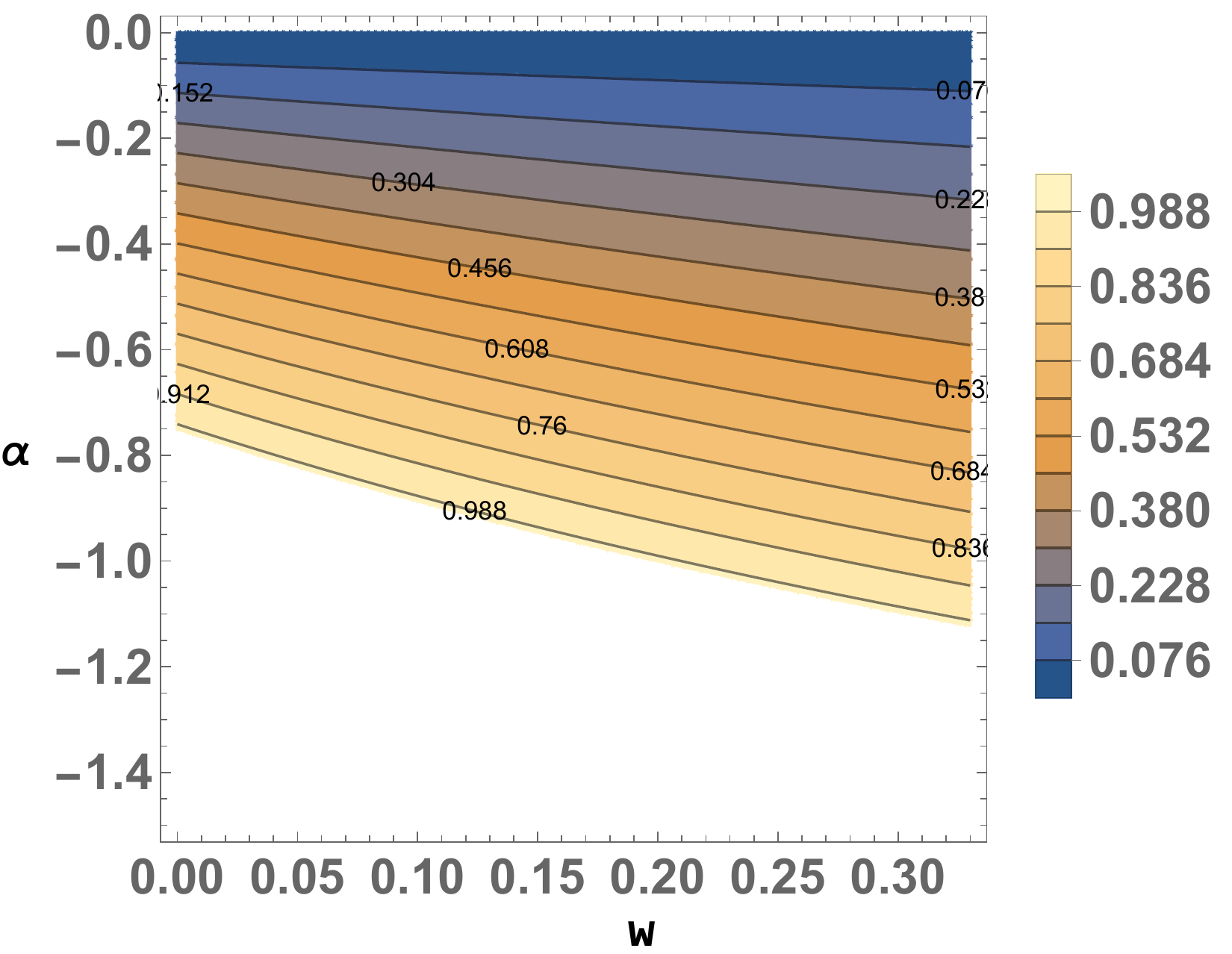}
\caption{The variation of the matter density parameter $s$ for the $A$ cosmological solution in a specific region of interest ($v=1, \beta=-1, x=1$).}
\label{fig:b1}       
\end{figure}

In terms of these auxiliary variables, we can write the Friedmann constraint equation \eqref{frcon} in the following way:
\begin{equation}
    s=\frac{1+u+v-x+y}{1+z},
\end{equation}
expressing the matter density parameter in terms of the remaining auxiliary variables.
\par 
Next, we introduce the e--fold number $N=log(a)$ and write the corresponding autonomous system of differential equations:
\begin{multline}
\frac{dx}{dN}=\frac{1}{4 \alpha  u z (z+1)} \Big[ 9 \beta  u^2 v w^2 y +12 \beta  u^2 v w y+3 \beta  u^2 v y+9 \beta  u v^2 w^2 y+12 \beta  u v^2 w y+3 \beta  u v^2 y -18 \beta  u v w^2 x^2+9 \beta  u v w^2 y^2
\\ +9 \beta  u v w^2 y-24 \beta  u v w x^2+12 \beta  u v w y^2 +12 \beta  u v w y-6 \beta  u v x^2+3 \beta  u v y^2+3 \beta  u v y+8 \alpha  u x z^2+8 \alpha  u x z
\\
-18 \beta  v^2 w^2 x^2+9 \beta  v^2 w^2 x y-24 \beta  v^2 w x^2+12 \beta  v^2 w x y-6 \beta  v^2 x^2 +3 \beta  v^2 x y+18 \beta  v w^2 x^3-18 \beta  v w^2 x^2-27 \beta  v w^2 x^2 y+9 \beta  v w^2 x y^2
\\
+9 \beta  v w^2 x y+24 \beta  v w x^3-24 \beta  v w x^2-36 \beta  v w x^2 y+12 \beta  v w x y^2 +12 \beta  v w x y+6 \beta  v x^3-6 \beta  v x^2-9 \beta  v x^2 y +3 \beta  v x y^2+3 \beta  v x y \Big],
\end{multline}

\begin{multline}
\frac{dy}{dN}=\frac{1}{4 \alpha  u z (z+1)} \Big[ -12 \alpha  u^2 w z-12 \alpha  u^2 z^2-12 \alpha  u^2 z -18 \beta  u v w^2 x-9 \beta  u v w^2 x y-24 \beta  u v w x-12 \beta  u v w x y
\\
-12 \alpha  u v w z -6 \beta  u v x-3 \beta  u v x y-12 \alpha  u v z^2-12 \alpha  u v z +12 \alpha  u w x z -12 \alpha  u w y z -12 \alpha  u w z+12 \alpha  u x z^2+12 \alpha  u x z
\\
-4 \alpha  u y z^2-4 \alpha  u y z-4 \alpha  u z^2 
-4 \alpha  u z-18 \beta  v^2 w^2 x -9 \beta  v^2 w^2 x y-24 \beta  v^2 w x-12 \beta  v^2 w x y-6 \beta  v^2 x-3 \beta  v^2 x y
\\
+18 \beta  v w^2 x^2+9 \beta  v w^2 x^2 y-18 \beta  v w^2 x-9 \beta  v w^2 x y^2-27 \beta  v w^2 x y +24 \beta  v w x^2+12 \beta  v w x^2 y-24 \beta  v w x-12 \beta  v w x y^2-36 \beta  v w x y
\\
+6 \beta  v x^2+3 \beta  v x^2 y-6 \beta  v x-3 \beta  v x y^2-9 \beta  v x y \Big],
\end{multline}

\begin{multline}
\frac{dz}{dN}=\Big[ -9 u w^3 z^2-9 u w^2 z^3-9 u w^2 z^2-3 u w z^2-3 u z^3 -3 u z^2-18 v w^3 z^2-9 v w^3 z-18 v w^2 z^3-39 v w^2 z^2
\\
-21 v w^2 z-12 v w z^3-30 v w z^2-15 v w z-6 v z^3 
-9 v z^2-3 v z+9 w^3 x z^2-9 w^3 y z^2-9 w^3 z^2
\\
+9 w^2 x z^3+9 w^2 x z^2-9 w^2 y z^3-9 w^2 y z^2-9 w^2 z^3 -9 w^2 z^2+3 w x z^2-3 w y z^2-3 w z^2+3 x z^3
\\
+3 x z^2-3 y z^3-3 y z^2-3 z^3-3 z^2 \Big] \cdot 
\\
\Big[ (3 w^2+1) z^2 (u-x+y+1)+v (w^2 (6 z^2+6 z+3) +4 w (2 z^2+3 z+1)+2 z^2+2 z+1) \Big]^{-1},
\end{multline}

\begin{multline}
\frac{du}{dN}=\frac{1}{4 \alpha  z (z+1)} \Big[ -18 \beta  u v w^2 x+9 \beta  u v w^2 y-24 \beta  u v w x 
+12 \beta  u v w y-6 \beta  u v x+3 \beta  u v y+8 \alpha  u z^2+8 \alpha  u z
\\
-18 \beta  v^2 w^2 x+9 \beta  v^2 w^2 y-24 \beta  v^2 w x+12 \beta  v^2 w y-6 \beta  v^2 x +3 \beta  v^2 y+18 \beta  v w^2 x^2-18 \beta  v w^2 x-27 \beta  v w^2 x y+9 \beta  v w^2 y^2
\\
+9 \beta  v w^2 y+24 \beta  v w x^2-24 \beta  v w x-36 \beta  v w x y+12 \beta  v w y^2 +12 \beta  v w y+6 \beta  v x^2-6 \beta  v x-9 \beta  v x y+3 \beta  v y^2+3 \beta  v y \Big],  
\end{multline}
\begin{multline}
 \frac{dv}{dN}=\Big[-v (u^2 (3 w^2+1) z^2 (3 \beta  v (w+1) (3 w+1) x  +2 \alpha  (z+1) (3 w+z+3))+u (3 \beta  v^2 (w+1) (3 w+1) x ((w (9 w+8)
 \\
 +3) z^2 +6 w (w+2) z+w (3 w+4)+2 z+1)  -2 v z (3 \beta  (w+1) (3 w+1) (3 w^2+1) x z (x-y-1)
 \\
 +\alpha  (z+1) (-9 w^3 z+3 w^2 (z^2+z+2)+w (z (16 z+21)+8)  +z^2+z+2))-2 \alpha  (3 w^2+1) z^2 (z+1) (3 w+z+3) (x-y-1))
 \\
 +3 \beta  v (w+1) (3 w+1) x (v-x+y+1) (v (w^2 (6 z (z+1)+3)  +4 w (z+1) (2 z+1)+2 z (z+1)+1)-(3 w^2+1) z^2 (x-y-1))) \Big] \cdot 
 \\
 \Big[ 2 \alpha  u z (z+1) ((3 w^2+1) z^2 (u-x+y+1)  +v (w^2 (6 z (z+1)+3)+4 w (z+1) (2 z+1)+2 z (z+1)+1)) \Big]^{-1}.
\end{multline}

\par 
In this case, where we have an exponential decomposition, we have obtained four critical points which are associated to a de--Sitter behavior ($w_{tot}=-1$). For these solutions the cosmological model acts as a geometrical dark energy component, driving the accelerated expansion of the Universe as a cosmological constant. In what follows, we shall describe each cosmological solution in detail, analyzing the dynamical consequences.
\par 
The first cosmological solution found for the exponential case is located at the following coordinates:
\begin{equation}
    A: \left( y= 0,z= -w-1,u= -\frac{3 \beta  v (3 w+1) x (v-x+1)}{9 \beta  v w x+3 \beta  v x-4 \alpha  w} \right),
\end{equation}
describing a de--Sitter epoch where the matter density parameter is equal to:
\begin{equation}
    s=-\frac{4 \alpha  (v-x+1)}{4 \alpha  w-3 \beta  v (3 w+1) x}.
\end{equation}
For this solution we note that the $x,z$ and $v$ variables are independent. The $y$ component which is related to the time variation of the Gauss--Bonnet geometrical invariant is set to zero. We can note that the location in the phase space structure is influenced by the barotropic equation of state of the matter sector, and $\alpha, \beta$ parameters which are describing the strength of the coupling functions. In Fig.~\ref{fig:b1} we have presented a possible region of interest for the matter density parameter where $s \in (0,1)$. In the general case where all the parameters are not set, the final expressions for the specific eigenvalues are too complex to be written in the manuscript. However, by setting some of the parameters ($\alpha=1, v=1, x=1, \beta=-1$), we have obtained some simple expressions of the resulting eigenvalues:  
\newpage 
\begin{multline}
    \Big[ 0,0,-\frac{3 (w+1) (w (51 w+22)+7)}{w (w (51 w+86)+19)+4},
    \\
    \frac{1}{2} \left(\frac{w (w+1) (3 w+1) (397 w+99) (w (w (51 w+86)+19)+4)}{\sqrt{w^2 (w+1)^2 (3 w+1)^2 (13 w+3) (397 w+99) (w (w (51 w+86)+19)+4)^2}}-3\right),
    \\
    \frac{1}{2}\left(-\frac{w (w+1) (3 w+1) (397 w+99) (w (w (51 w+86)+19)+4)}{\sqrt{w^2 (w+1)^2 (3 w+1)^2 (13 w+3) (397 w+99) (w (w (51 w+86)+19)+4)^2}}-3\right) \Big].
\end{multline}
\par
\begin{figure}
  \includegraphics[height=4cm]{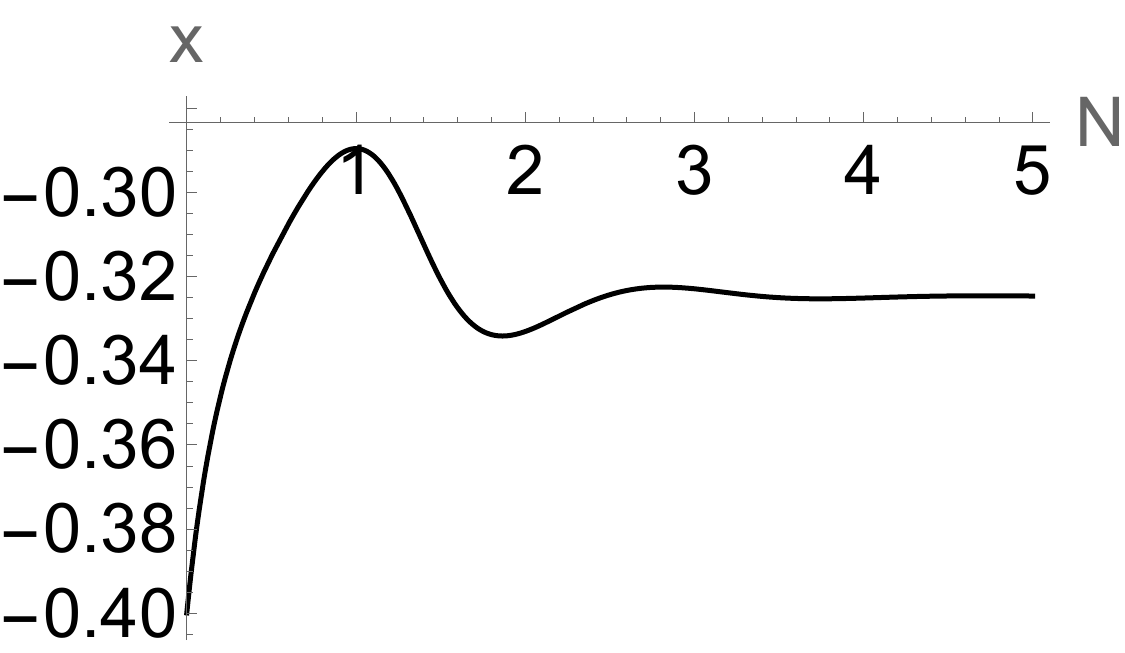}
  \includegraphics[height=4cm]{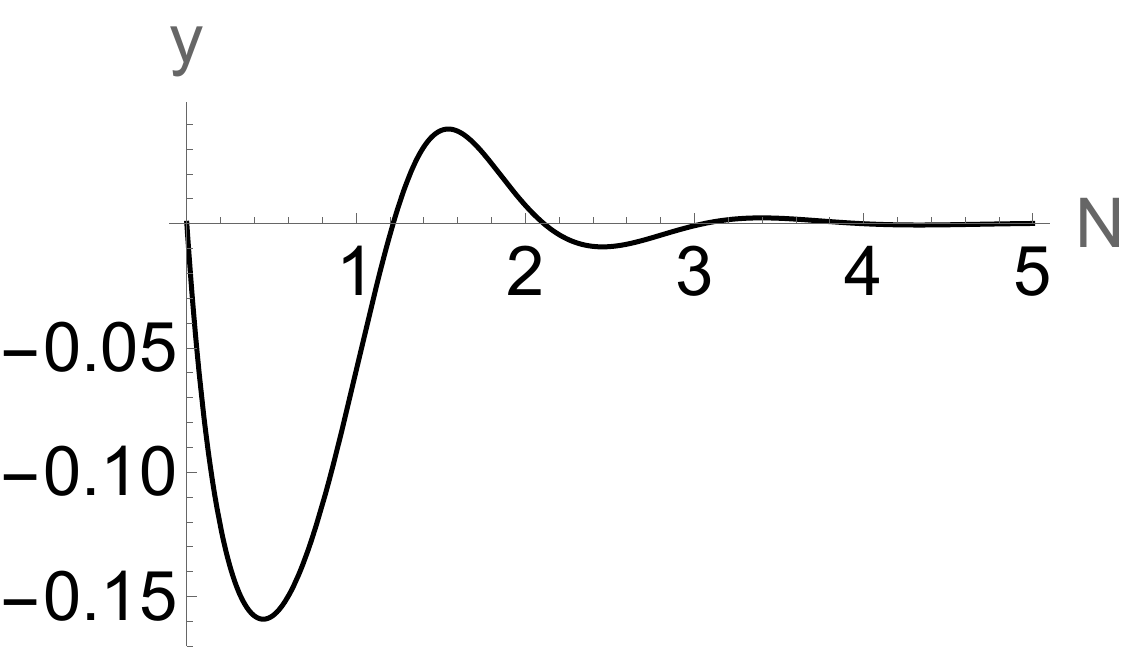}
  \includegraphics[height=4cm]{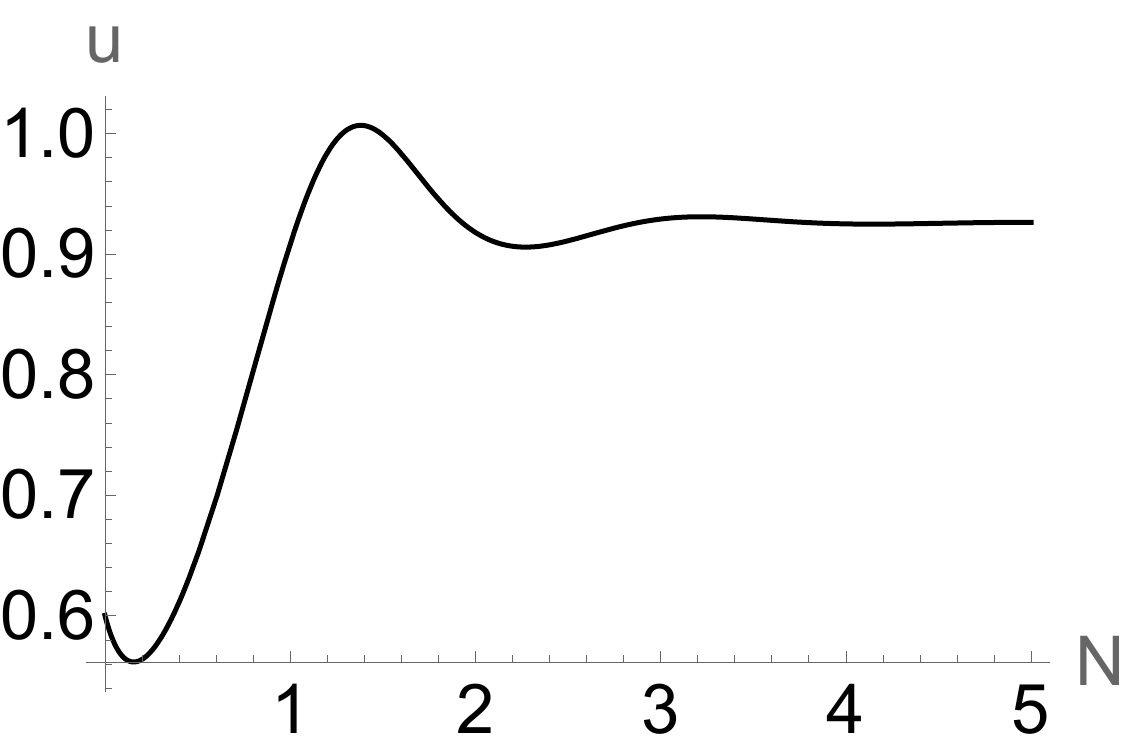}
  \includegraphics[height=4cm]{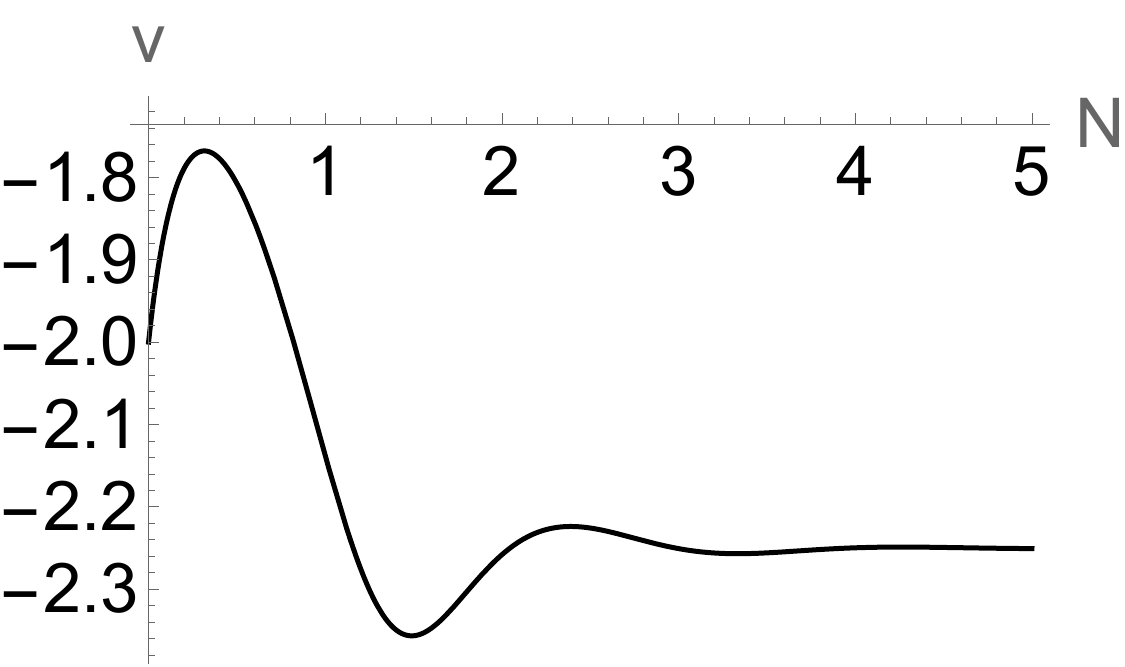}
\caption{The evolution towards the A cosmological solution in the phase space structure ($w=0.0001, \alpha=-2, \beta=-1$).}
\label{fig:punctA1}       
\end{figure}

\begin{figure}[tb]
  \includegraphics[width=6cm]{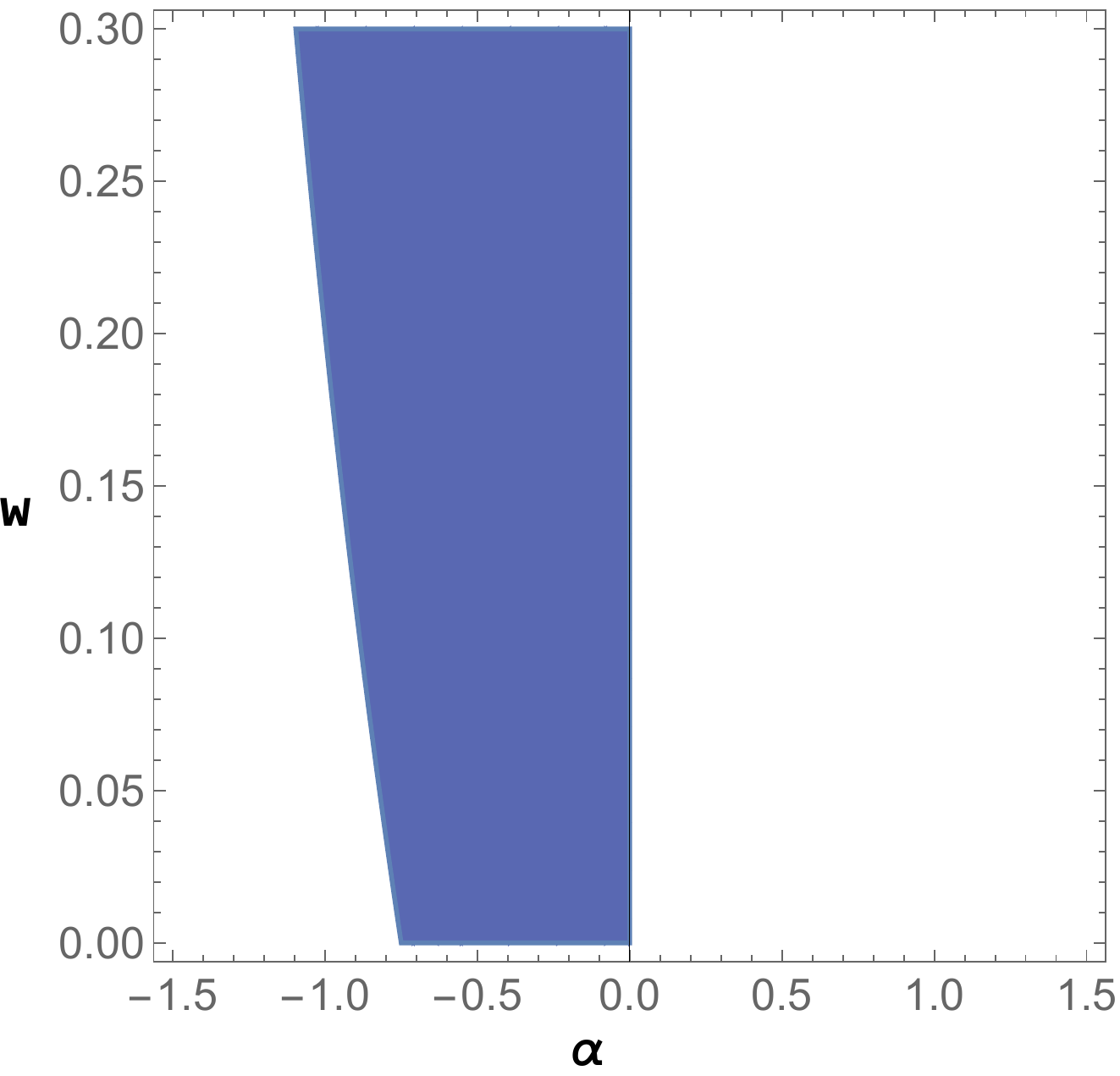}
\caption{A specific region of interest for the $A$ cosmological solution where the dynamics corresponds to a saddle dynamical behavior. ($v=1, \beta=-1, x=1$).}
\label{fig:b2}       
\end{figure}

 
As can be observed, the solution is non--hyperbolic, due to the existence of two zero eigenvalues. Hence, we can use the linear stability theory only to study the specific cases where the dynamics corresponds to a saddle behavior. A specific region where we have obtained a saddle behavior is presented in Fig.~\ref{fig:b2}. The evolution in the phase space structure towards the A cosmological solution can be seen in Fig.~\ref{fig:punctA1}. In this case the corresponding eigenvalues have the following values:
\begin{equation}
    \big[ 0.,0.,-1.00194,-1.5+3.36584 i,-1.5-3.36584 i\big].
\end{equation}

\par
\begin{figure}[t]
  \includegraphics[width=6cm]{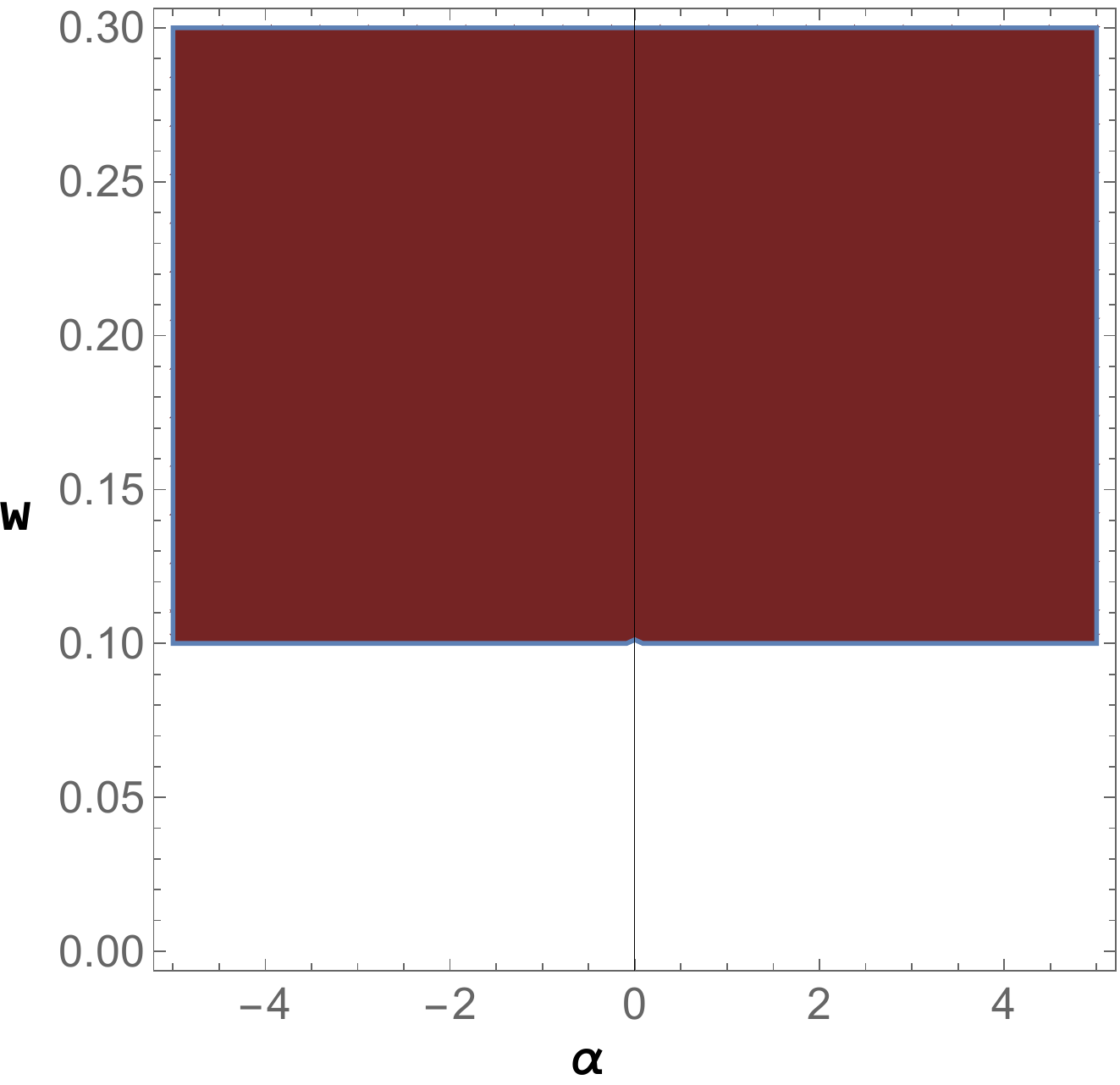}
\caption{A specific region of interest for the $B^{+}$ cosmological solution where the dynamics corresponds to a saddle dynamical behavior. ($u=-0.1, \beta=1$).}
\label{fig:b3}       
\end{figure}

 
The second cosmological solution is found at the coordinates:
\begin{multline}
    B^{\pm}: \Big( x= \frac{3 \beta \pm \sqrt{3} \sqrt{\beta  (3 w+1) (3 \beta +16 \alpha  w+9 \beta  w)}+9 \beta  w}{6 \beta +18 \beta  w}, y= 0,z= -w-1,
    \\
    v= \frac{\pm \sqrt{3} \sqrt{\beta  (3 w+1) (3 \beta +16 \alpha  w+9 \beta  w)}-3 \beta  (3 w+1)}{6 (\beta +3 \beta  w)} \Big),
\end{multline}
with the corresponding matter density parameter equal to: $s=-\frac{u}{w}$. Due to the specific form of the matter density parameter, we can observe that the case of a pressure--less dark matter fluid cannot be considered, leading to a divergence. Hence, we can only approximate the case of a pressure--less dark matter fluid, $w \to 0$. The eigenvalues for the $B^{+}$ solutions are the following:
\begin{multline}
    \Big[ 0,0, -\frac{3 (w+1) (3 w+1) \left(6 \beta  u+9 \beta  (2 u-1) w^2+\sqrt{3} w \sqrt{\beta  (3 w+1) (3 \beta +16 \alpha  w+9 \beta  w)}-3 \beta  w\right)}{6 \beta  u (w+1) (3 w+1) \left(3 w^2+1\right)+(3 w+5) w^2 \left(\sqrt{3} \sqrt{\beta  (3 w+1) (3 \beta +16 \alpha  w+9 \beta  w)}-3 \beta  (3 w+1)\right)}, E_4, E_5
     \Big], 
\end{multline}
where $E_4, E_5$ have complicated expressions and are not displayed in the manuscript. In Fig.~\ref{fig:b3} we have displayed a saddle region for the $B^{+}$ solution, taking into account also the existence conditions which imply $s \in (0,1)$.

\par 
The last de--Sitter cosmological solution found in the present analysis is located in the phase space structure at the coordinates:
\begin{multline}
   C: \Big( x = \frac{4 \left(\alpha +\alpha  v+6 \alpha  v w^2+\alpha  v w+3 \alpha  w^2\right)}{4 \alpha +3 \beta  v^2+27 \beta  v^2 w^2+18 \beta  v^2 w+12 \alpha  w^2}, y= 0,z= -w-1, u=
   \\
   -\frac{3 \beta  v^2 (3 w+1)^2 \left(6 v w^2+v w+v+3 w^2+1\right)}{\left(3 w^2+1\right) \left(4 \alpha +3 \beta  v^2+27 \beta  v^2 w^2+18 \beta  v^2 w+12 \alpha  w^2\right)} \Big),
\end{multline}
with the matter density parameter influenced by the dark matter equation of state and specific coupling of the energy--momentum--squared function,
\begin{equation}
    s=\frac{v (3 w+1)}{3 w^2+1}.
\end{equation}
\par
\begin{figure}[t]
  \includegraphics[width=6cm]{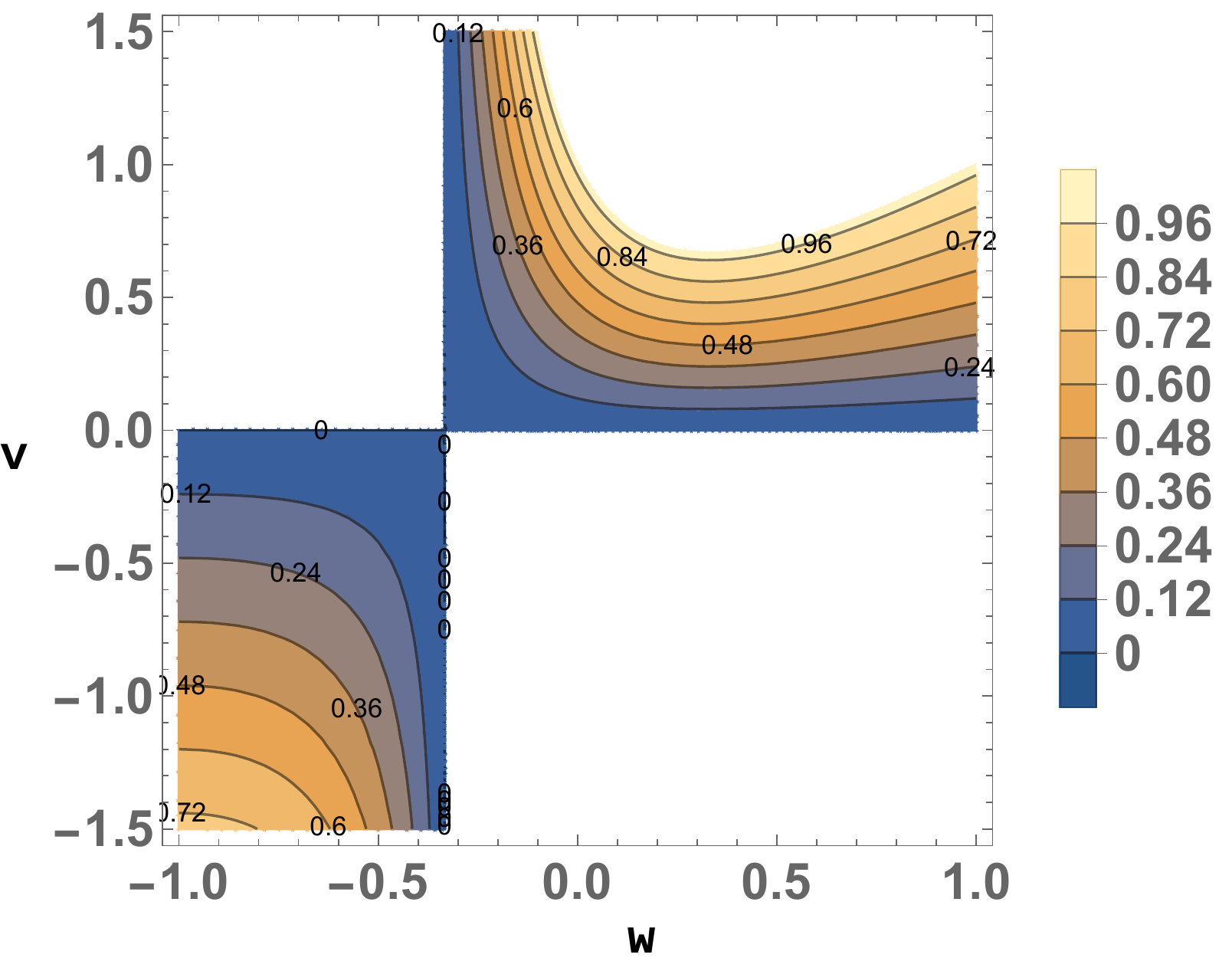}
\caption{The variation of the matter density parameter $s$ for the $C$ cosmological solution.}
\label{fig:b4}       
\end{figure}

 
The variation of the corresponding matter density parameter for the $C$ cosmological solution is presented in Fig.~\ref{fig:b4}. It can be seen that in the case of a pressure--less dark matter component the matter density parameter satisfies the observational constraints, being influenced also by the $v$ variable which encodes geometrical effects due to the specific form of the energy--momentum--squared function.
\par 
For this critical line we have obtained the following eigenvalues ($v=1,\beta=1$):
\begin{equation}
    \Bigg[ 0,0, 0, \frac{C_1 \pm\frac{\sqrt{2} \sqrt{w^2 (3 w+1)^4 \left(w^2-1\right)^2 \left(3 w^2+1\right) \left(9 w^2+w+2\right) \left(50 \alpha ^2 \left(3 w^2+1\right) \left(9 w^2+w+2\right)+12 \alpha  \left(3 w^2+1\right) (3 w+1)^2+9 (3 w+1)^4\right) (4 \alpha +3 w ((4 \alpha +9) w+6)+3)^2}}{\alpha  (w-1) w (w+1) (3 w+1)^2 \left(3 w^2+1\right) \left(9 w^2+w+2\right)}}{4 (4 \alpha +3 w ((4 \alpha +9) w+6)+3)}
     \Bigg], 
\end{equation}
where we have defined:
\begin{equation}
    C_1=6 (-4 \alpha -3 w ((4 \alpha +9) w+6)-3).
\end{equation}
\par
\begin{figure}[t]
  \includegraphics[width=6cm]{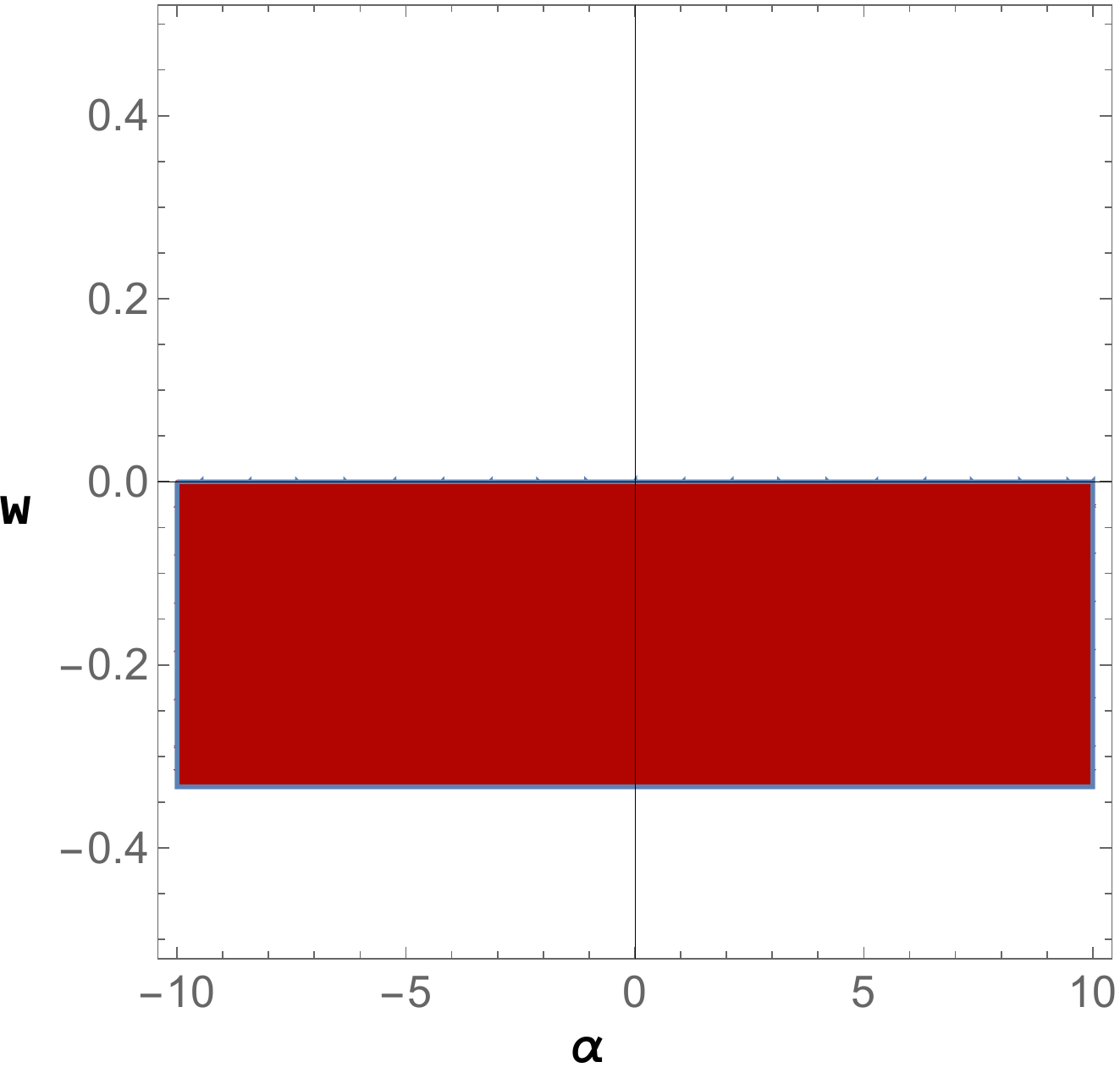}
\caption{A specific region of interest for the $C$ cosmological solution where the dynamics corresponds to a saddle dynamical behavior. ($v=1, \beta=1$).}
\label{fig:b5}       
\end{figure}

\begin{figure}
  \includegraphics[width=6cm]{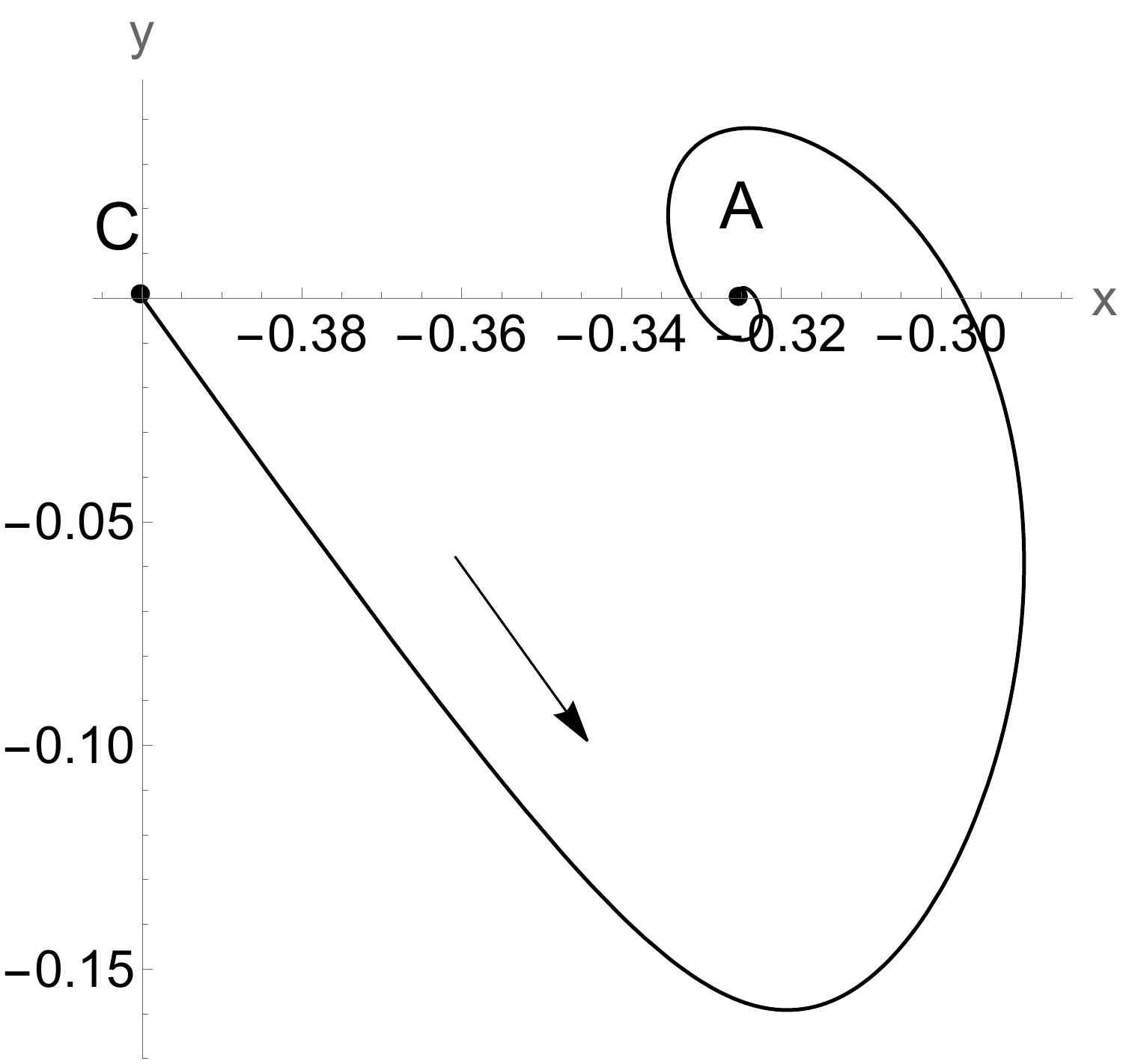}
\caption{The evolution from the critical point C towards the A cosmological solution in the xOy plane where $w=0.0001, \alpha=-2, \beta=-1$.}
\label{fig:tranzitie}       
\end{figure}

\begin{figure}
  \includegraphics[width=6cm]{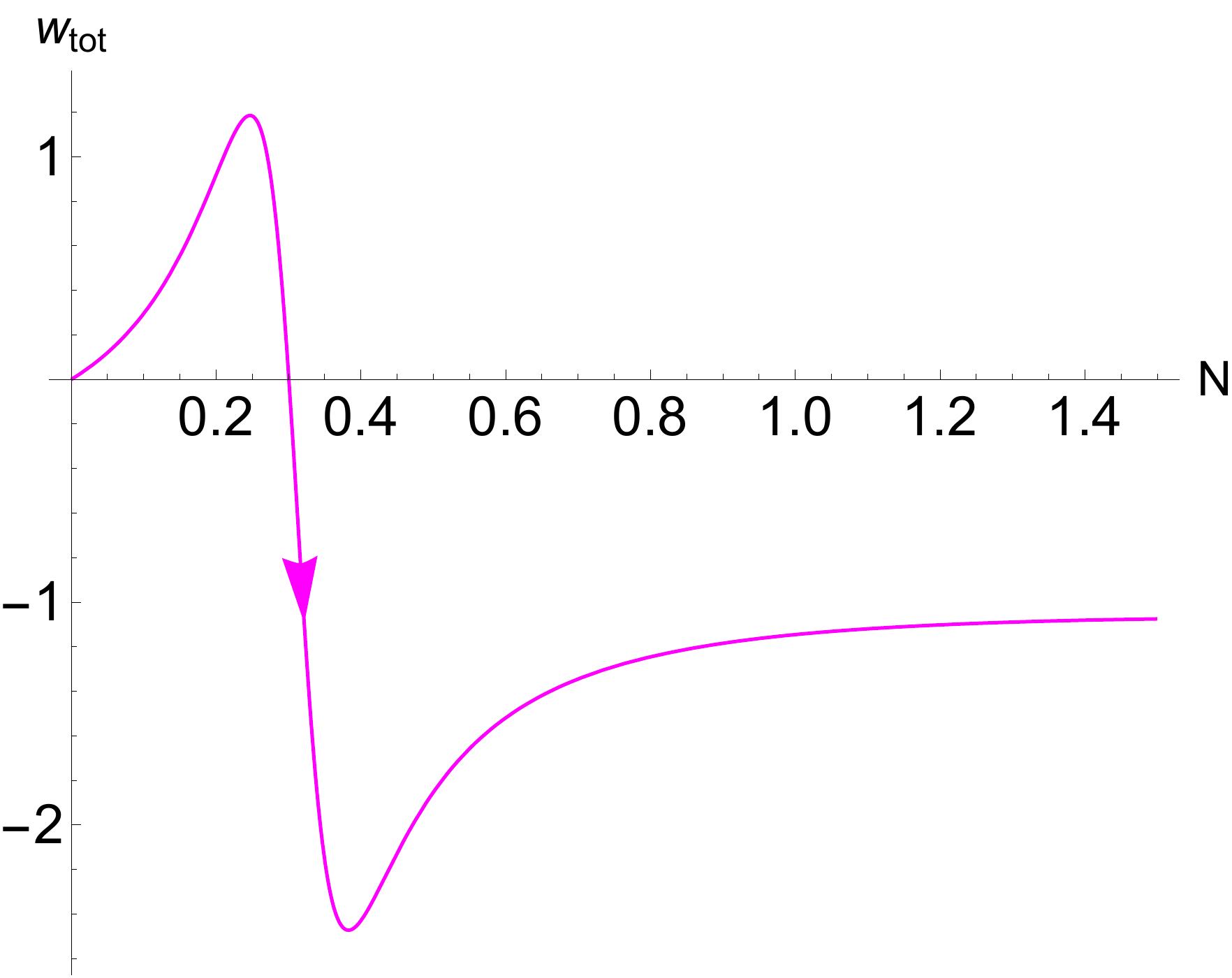}
\caption{The variation of the total equation of state in the case where $w=0.0001, \alpha=-2, \beta=-1$.}
\label{fig:wtotal}       
\end{figure}

 
In Fig.~\ref{fig:b5} we have presented a region where the dynamical corresponds to a saddle behavior, possible explaining the late time acceleration of the Universe in the background dynamics. Note that the last critical line is also non--hyperbolic, having three zero eigenvalues. The transition from the critical point $C$ towards the $A$ cosmological solution in the xOy plane can be observed in Fig.~\ref{fig:tranzitie} for specific initial conditions near the $C$ solution, validating the obtained analytical solutions. Lastly, in Fig.~\ref{fig:wtotal} we have represented the variation of the total (effective) equation of state for the present cosmological scenario. We note that the evolution can pass from a matter domination epoch towards a super--accelerated era, attaining the cosmological constant boundary from below at late times, crossing the phantom divide line in the early stages.

\section{Summary and Conclusions}
\label{sec:3}
\par 

In this paper we have proposed a model in the theoretical framework of modified gravity, where the fundamental Einstein--Hilbert action is extended, by considering a more complete theory. The latter theory denoted as $F(G,T_{\mu \nu}T^{\mu \nu})$ is based on two specific components. The first component takes into account possible physical effects due to the consideration of the Gauss--Bonnet invariant $(G)$, encoding geometrical aspects in the generic theory. The second component in our action is based on the energy--momentum--squared invariant $(T_{\mu \nu}T^{\mu \nu})$, embedding geometrical effects from the specific form of the energy--momentum tensor. In this cosmological model we have assumed that the generic function which depends on the Gauss--Bonnet invariant and the energy--momentum--squared term can be decomposed in an independent manner, $F(G, T_{\mu\nu}T^{\mu\nu})=f(G)+g(T_{\mu\nu}T^{\mu\nu})$.  After we have proposed the generic action for our present model, we have obtained the modified Friedmann relations by varying the action with respect to the inverse metric, assuming that the background can be described by the Robertson--Walker metric. Here we note that the continuity equation is not satisfied due to the inclusion of the energy--momentum--squared term in the specific action, a particular aspect for these theories. After obtaining the dynamical equations, we have studied the physical aspects of our cosmological model by considering the linear stability theory. In this study we have assumed an exponential representation for the generic function in our action, $F(G, T_{\mu\nu}T^{\mu\nu})=f_0 e^{\alpha G}+g_0 e^{\beta T^2}$, where $\alpha, \beta, f_0, g_0$ are constant parameters. In this particular case we have introduced the auxiliary variables which are required in order to apply the linear stability theory. After introducing the auxiliary variables associated to the phase space structure, we have computed the critical points of the present cosmological model for the exponential case. 
\par 
In the phase space structure we have identified various  critical points which correspond to a de--Sitter epoch where the model behaves closely as a cosmological constant, particular solutions which can explain the late time stage of the Universe. As can be seen from the analysis, for these solutions the effective matter density parameter is influenced by different coupling terms, and the barotropic equation of state for the matter sector. From a dynamical point of view we have identified possible regions of interest for the coupling constants which correspond to a saddle dynamical behavior in the late time stage of the Universe. These critical points are particular solutions which correspond in principle to various epochs in the dynamical trajectory of our Universe. In this case the phase space structure has four critical points, associated to a de--Sitter epoch. For each of the corresponding critical points we have established the dynamical behavior, obtaining possible regions of interest for different parameters which can describe the present model. 
\par 
The analysis of the phase space structure showed that the present cosmological model can describe the late--time accelerated expansion of the Universe and the dynamical behavior of the effective equation of state. However, due to the de--Sitter solutions found in the phase space structure, we have to further assume that the matter and the radiation epochs appear by fine--tuning the initial conditions of the current trajectory. The present paper can be extended in various cosmological applications. For example, it would be interesting to study the cosmological model by considering an observational study with the recent cosmological data, obtaining constraints for various parameters from an astrophysical point of view. Another possible aspect is represented by the inflationary era, a particular stage in the evolution of the Universe which can be further analyzed. These particular extensions can provide support for a more complete theory of gravity and are left as future projects.

\begin{acknowledgments}
\label{sec:5}
We would like to thank Prof. Dr. Virgil Baran for various discussions which lead to the development of the present project. The computational part of this work was performed using the computer stations provided by CNFIS through the project CNFIS-FDI-2020-035.
\end{acknowledgments}

\bibliography{sorsamp}

\end{document}